\documentclass[journal]{IEEEtran}

\usepackage{amsmath,amssymb,mathtools}
\usepackage{graphicx}
\usepackage{booktabs}
\usepackage{array}
\usepackage{tabularx}
\usepackage{siunitx}
\usepackage{cite}
\usepackage{microtype}
\usepackage{xspace}
\usepackage{hyperref}
\hypersetup{
  pdftitle={A Givens-Exchange Ansatz for Molecular Variational Eigensolvers},
  pdfauthor={Azadeh Alavi, Fatemeh Kouchmeshki, Muhammad Usman, Yongli Ren, Ke Deng, Hossein Akhoundi, and Abdolrahman Alavi},
  pdfsubject={Fixed-topology Givens-exchange ansatz design and benchmarking for molecular variational quantum eigensolvers},
  pdfkeywords={variational quantum eigensolver, ansatz design, Givens rotations, quantum architecture search, molecular simulation}
}

\newcommand{\mHa}{\mathrm{mHa}}
\newcommand{\Ha}{\mathrm{Ha}}
\newcommand{\JW}{Jordan--Wigner\xspace}

\newcommand{\HtwoO}{H$_2$O\xspace}
\newcommand{\BeHtwo}{BeH$_2$\xspace}
\newcolumntype{Y}{>{\centering\arraybackslash}X}

\title{A Givens-Exchange Ansatz for Molecular Variational Eigensolvers}

\author{Azadeh~Alavi, Fatemeh~Kouchmeshki, Muhammad~Usman, Yongli~Ren, Ke~Deng, Hossein~Akhoundi, and Abdolrahman~Alavi%
\thanks{Azadeh Alavi and Fatemeh Kouchmeshki contributed equally to this work. This work received no external funding.}%
\thanks{Azadeh Alavi, Yongli Ren, and Ke Deng are with the School of Computing Technologies, RMIT University, Melbourne, VIC, Australia.}%
\thanks{Fatemeh Kouchmeshki, Hossein Akhoundi, and Abdolrahman Alavi are with Pattern Recognition Pty Ltd, Australia.}%
\thanks{Muhammad Usman is with Quantum Systems, CSIRO, Clayton, VIC 3168, Australia. and with School of Physics, The University of Melbourne, Parkville, 3010 Victoria Australia
}%
\thanks{Corresponding authors: Azadeh Alavi (azadeh.alavi@rmit.edu.au) and Fatemeh Kouchmeshki (admin@pr2aid.com).}}

\begin{document}
\maketitle

\begin{abstract}
Ansatz design in variational quantum eigensolvers (VQEs) requires a balance among expressivity, trainability, and implementation cost. Quantum architecture search can identify compact circuits, but it adds an outer circuit-construction loop and associated computational overhead. We develop and evaluate a deterministic, fixed-topology Givens-exchange ansatz for molecular VQE. The circuit initializes the computational-basis state with the lowest diagonal Hamiltonian expectation and applies three local single-qubit rotation layers interleaved with two ordered all-pair Givens-exchange blocks. Parameters are optimized solely from Hamiltonian expectation values; exact diagonalization is reserved for post-optimization errors and fidelities. Across six predetermined seeds, every run on coefficient-verified LiH-6 and H$_2$O-8 Hamiltonians and on a BeH$_2$-6 public-specification candidate satisfies chemical accuracy. The respective six-seed mean errors are $0.000124$, $0.128558$, and $0.002152\,\mHa$. For LiH-6 and H$_2$O-8, these means are lower than the published point errors of the compared quantum-architecture-search methods, although the present ansatz uses a larger pre-compilation macro budget. Resource quantities are reported at their native levels, and hardware realization is discussed in terms of native-gate compilation, connectivity, routing, finite-shot estimation, and noisy reoptimization. The method provides an accurate, transparent, and search-free reference architecture for molecular VQE engineering.
\end{abstract}


\section{Introduction}
Ansatz design is a central engineering problem in the variational quantum eigensolver (VQE). A useful circuit must be expressive enough to represent correlated ground states, trainable under a finite optimization budget, and realizable with acceptable native-gate depth, connectivity overhead, and measurement cost. These objectives are coupled: shallow hardware-efficient circuits may omit molecular excitation structure, whereas chemistry-informed ansatzes can become expensive after compilation. Automated quantum architecture search (QAS) treats circuit topology as an optimization variable, but introduces an additional circuit-construction and search loop before the final variational parameters are selected.

VQE estimates a ground-state energy by minimizing the expectation value of a qubit Hamiltonian over a parameterized quantum state \cite{peruzzo2014vqe,mcclean2016theory,cerezo2021vqa}. Molecular ground-state energies support conformer ranking, reaction and binding energetics, and electronic-structure analysis, including workflows in computational drug discovery \cite{cao2019,blunt2022,mcardle2020}. Bond breaking, transition-metal chemistry in metalloenzymes, and other strongly correlated active spaces are especially challenging because classical electronic-structure approximations can become expensive or unreliable. Recent noisy-hardware studies have extended variational and related electronic-structure workflows through symmetry checking, machine-learning-assisted VQE optimization, full band-structure calculations, and low-depth virtual distillation \cite{zhang2025symmetry,karim2025mlvqe,zhang2024band,karim2024virtual}. The ansatz trade-off is therefore consequential both for numerical accuracy and for implementation on near-term quantum hardware.

QAS addresses this trade-off by automating gate selection and placement. Reinforcement learning, evolutionary search, and adaptive pool-based methods have been used to choose gates, entangling patterns, and parameters for variational circuits \cite{adaptvqe,sim2019,hyrlqas}. Hybrid-Action Reinforcement Learning for Quantum Architecture Search (HyRLQAS) combines discrete gate choices with continuous parameter information and reports strong noiseless VQE results on LiH, \BeHtwo, and \HtwoO benchmarks \cite{hyrlqas}. Its comparison set includes TensorRL-QAS, curriculum reinforcement-learning QAS (CRLQAS), Vanilla reinforcement learning, BenchRL-QAS variants, quantumDARTS, and training-free QAS \cite{tensorrl,crlqas,vanillarl,benchrlqas,quantumdarts,tfqas,hyrlqas}. Search-based methods can discover compact circuits, but generally require repeated circuit construction, parameter refinement, and policy or search updates before a final circuit is retained \cite{hyrlqas,crlqas,benchrlqas}.

Here, we evaluate a fixed-topology Givens-exchange ansatz \cite{kivlichan2018,xia2020qccsd,gard2020,anselmetti2021,arrazola2022,nirmal2025givens} with a Stiefel-coordinate interpretation \cite{edelman1998,absil2008,smart2024manifold,qin2026geometric}. The Stiefel manifold supplies geometric language for the unitary evolution of normalized states and orthonormal frames, while Givens-exchange macros provide a chemically motivated two-qubit operation that redistributes single-excitation amplitudes in occupation encodings. The evaluated protocol combines two ordered all-pair exchange blocks, three local $R_Y$ layers, diagonal-basis initialization, and Adam optimization of the Hamiltonian expectation value.

The contributions are as follows:
\begin{itemize}
    \item Building on prior Givens-rotation and symmetry-preserving exchange-circuit constructions \cite{kivlichan2018,xia2020qccsd,gard2020,anselmetti2021,arrazola2022,nirmal2025givens}, we define the specific fixed-topology Givens-exchange ansatz studied here and state the ordered sequence of rotation and exchange layers used in every reported evaluation.
    \item We specify the optimization and evaluation information flow: variational updates use only the Hamiltonian expectation objective, while exact diagonalization is introduced only after optimization to compute errors and fidelities.
    \item We separate the primary \JW comparisons from the parity-mapped and tapered LiH-4 stress case, and verify coefficient identity for the LiH-6 and H$_2$O-8 primary Hamiltonians against the corresponding reference benchmarks.
    \item We report six-seed means, sample standard deviations, medians, extrema, chemical-accuracy pass counts, fidelities, and wall times rather than selecting a single favorable run.
    \item We report tabulated energy errors in mHa, with the exact Ha--mHa conversion stated in the mathematical formulation, and compare the fixed topology with published QAS baselines without equating Givens macros with native CNOT counts.
    \item We distinguish the absence of an architecture-search loop from hardware-native circuit cost and state the compilation, routing, finite-shot, and noisy-reoptimization steps required for hardware assessment.
\end{itemize}

\section{Background and related work}
\label{sec:background}

Molecular variational quantum eigensolvers (VQE) begin from the electronic-structure Hamiltonian at a fixed nuclear geometry and in a chosen one-particle basis. In second quantization, this Hamiltonian is commonly written as
\begin{equation}
    \hat{H}_{\mathrm{el}}
    =
    \sum_{pq} h_{pq} a_p^\dagger a_q
    +
    \frac{1}{2}
    \sum_{pqrs} h_{pqrs}
    a_p^\dagger a_q^\dagger a_r a_s ,
    \label{eq:second_quantized_background}
\end{equation}
where the indices $p,q,r,s$ run over the spin orbitals of the chosen basis. The operators $a_p^\dagger$ and $a_p$ respectively create and annihilate an electron in spin orbital $p$; analogously, $a_r$ and $a_s$ annihilate electrons in spin orbitals $r$ and $s$ in the two-electron term. The coefficients $h_{pq}$ and $h_{pqrs}$ are the one- and two-electron molecular integrals, encoding the kinetic, electron--nuclear, and electron--electron contributions to the electronic energy. The compact form of Eq.~\eqref{eq:second_quantized_background} relies on the canonical anticommutation relations of the fermionic creation and annihilation operators, which enforce the antisymmetry of electronic wavefunctions. Since a quantum processor acts on qubits through Pauli operators, the fermionic Hamiltonian must be transformed into an equivalent qubit Hamiltonian whose Pauli terms can be measured individually \cite{whitfield2011,cao2019,mcardle2020}.

This transformation is carried out by a fermion-to-qubit mapping, which replaces products of fermionic operators by tensor products of Pauli operators while preserving the sign structure induced by fermionic exchange. The Jordan--Wigner (JW) transformation encodes the occupation of each spin orbital directly in a single qubit and represents fermionic parity signs through strings of Pauli-$Z$ operators \cite{jordan1928,whitfield2011}. This encoding gives a transparent occupation-number interpretation, although the resulting Pauli strings may grow linearly with the number of spin orbitals. The parity mapping instead stores cumulative occupation parities, making molecular $\mathbb{Z}_2$ symmetries more explicit and naturally supporting qubit tapering. The Bravyi--Kitaev (BK) transformation balances occupation and parity information, reducing the asymptotic length of the resulting Pauli strings relative to JW in many cases \cite{bravyi2002,seeley2012,tranter2018}. When symmetry sectors are known, tapering fixes selected qubit degrees of freedom to prescribed symmetry eigenvalues and thereby reduces the number of active qubits without changing the targeted physical sector \cite{bravyi2017tapering}. These mapping choices are consequential for ansatz design: an operation that has an immediate particle-conserving interpretation in a JW occupation basis may become a more abstract effective variational operation after parity mapping or tapering.

After mapping, VQE applies the variational principle to the qubit Hamiltonian. A parameterized circuit prepares a trial state $|\psi(\theta)\rangle$, the quantum device or simulator estimates the expectation value $\langle\psi(\theta)|H|\psi(\theta)\rangle$, and a classical optimizer updates the parameters $\theta$ to minimize it \cite{peruzzo2014vqe,mcclean2016theory,cerezo2021vqa}. The energy, shift, error, and fidelity definitions used in this work are stated formally in Section~\ref{sec:formulation}. The central modelling decision is the choice of ansatz. Chemistry-inspired ansatzes, such as unitary coupled-cluster variants including UCCSD, encode fermionic excitation structure and are well aligned with molecular electronic correlation, but their compiled circuits can be too deep for near-term hardware \cite{romero2018ucc,mcardle2020}. Hardware-efficient ansatzes instead use shallow layers of native one- and two-qubit gates, reducing circuit depth but often sacrificing particle-number, spin, and excitation-structure constraints; in sufficiently large Hilbert spaces, they can also suffer from poor trainability \cite{sim2019,cerezo2021vqa}. Adaptive approaches such as ADAPT-VQE construct circuits incrementally from an operator pool, selecting each new excitation according to energy-gradient information and thereby improving compactness at the cost of repeated operator screening \cite{adaptvqe}. Designing ansatzes that are simultaneously accurate, compact, interpretable, and trainable therefore remains a central challenge in molecular VQE.

A complementary line of work seeks to automate ansatz construction. Quantum architecture search (QAS) treats the circuit architecture itself as an optimization variable, using evolutionary strategies, gradient-based methods, and reinforcement learning to balance expressivity, trainability, hardware cost, and noise sensitivity \cite{du2022qas,wang2022quantumnas}. Reinforcement-learning formulations have been used to optimize gate placement and circuit structure for VQE problems, and HyRLQAS extends this direction through a hybrid action space that combines discrete gate choices with continuous parameter information \cite{hyrlqas}. Although such methods can identify compact and high-performing circuits, the resulting architectures may be difficult to interpret chemically, and the search process can introduce substantial computational overhead. These limitations motivate ansatz families that retain chemically meaningful structure while avoiding both fully hand-designed excitation expansions and unrestricted learned architecture search.

Givens rotations provide one such structured prior. In numerical linear algebra, a Givens rotation acts within a two-dimensional coordinate plane while leaving all other coordinates unchanged. In occupation-based encodings for quantum chemistry, the analogous two-qubit single-excitation operation rotates the subspace spanned by $|01\rangle$ and $|10\rangle$ while leaving $|00\rangle$ and $|11\rangle$ invariant. This structure aligns Givens-type gates with orbital rotations, Slater-determinant preparation, and particle-conserving circuit families. Particle-preserving exchange gates have been used to realize qubit excitations, symmetry-preserving circuits have been constructed to span prescribed particle-number and spin sectors, and local quantum-number-preserving fabrics combine orbital rotations with pair-exchange gates \cite{xia2020qccsd,gard2020,anselmetti2021}. Fermionic swap networks and Givens rotations have also been used to derive low-depth schemes for electronic-structure simulation and Slater-determinant preparation \cite{kivlichan2018}. Subsequent work has shown that controlled single-excitation Givens rotations can support universal particle-conserving quantum-chemistry circuits, and a parallelized Givens ansatz has been proposed as a low-depth alternative to deeper coupled-cluster-style circuits for molecular VQE \cite{arrazola2022,nirmal2025givens}. These studies establish that neither the Givens rotation nor the exchange primitive is new here. Their objectives, however, differ from the present benchmarking question: symmetry-preserving constructions constrain evolution to selected quantum-number sectors, whereas the parallelized Givens construction targets depth reduction relative to UCCSD using its own active-space and circuit-layout choices. The circuit evaluated here deliberately interleaves exchange blocks with local $R_Y$ rotations and therefore does not preserve particle number; it instead tests one deterministic lexicographic all-pair template against coefficient-verified QAS benchmarks. The contribution is consequently the explicit protocol and controlled search-free benchmark, rather than a new exchange primitive or a universal symmetry-preserving construction.

The geometric viewpoint adopted in this manuscript is expressed through the complex Stiefel manifold $\mathrm{St}_{\mathbb{C}}(D,r)$, the space of orthonormal $r$-frames in $\mathbb{C}^{D}$ \cite{edelman1998,absil2008}; its definition and the norm-preserving update used here are given in Section~\ref{sec:construction}, Eq.~\eqref{eq:stiefel}. Earlier quantum-chemistry geometry work established Stiefel and Grassmann structures for Slater-type variational spaces in Hartree--Fock theory and beyond \cite{chiumiento2012}. Its objects are orthonormal one-particle orbital frames and subspaces, rather than finite-depth parameterized qubit circuits, so it does not specify the ordered exchange topology or the VQE benchmark protocol studied here. A normalized $n$-qubit pure-state representative is a point on $\mathrm{St}_{\mathbb{C}}(2^n,1)$, with physically equivalent representatives related by a global phase, and the first $r$ columns of an $n$-qubit unitary form a point on $\mathrm{St}_{\mathbb{C}}(2^n,r)$. Direct quantum-manifold approaches instead minimize the many-body eigenstate problem over Stiefel or Grassmann manifolds to enforce orthogonality and obtain multiple eigenstates without an explicit state parameterization, while recent ansatz-free geometric analyses study VQE landscapes, initialization, and Riemannian convergence on unitary manifolds \cite{smart2024manifold,qin2026geometric}. Those approaches provide optimization frameworks or theoretical guarantees, but they do not define or benchmark the finite-depth Givens--$R_Y$ topology used here. Conversely, the present work does not perform Riemannian optimization and does not inherit their convergence or multi-state guarantees; the Stiefel language is used only to describe the normalized unitary trajectory of the chosen circuit.

The Givens-exchange ansatz studied in this work lies at the intersection of these developments. It uses the qubit Hamiltonian obtained from standard fermion-to-qubit mappings, retains the conventional VQE energy objective, and evaluates a fixed all-pair exchange template as a geometry-guided and chemically interpretable molecular prior. The benchmark design follows the molecular labels used in the HyRLQAS comparison while explicitly distinguishing the LiH-6 Jordan--Wigner comparison from the parity-mapped LiH-4 stress case. For the primary LiH-6 and \HtwoO-8 benchmark rows, supplementary verification confirms consistency between the Pauli-coefficient representation used in our experiments and the corresponding reference Hamiltonians. This positioning allows us to assess whether a transparent Givens-exchange prior can provide competitive molecular-VQE performance without relying on unrestricted learned architecture search.

\section{Mathematical formulation and evaluation protocol}
\label{sec:formulation}
Each molecular problem is represented after fermion-to-qubit mapping by an $n$-qubit Hamiltonian,
\begin{equation}
    H=\sum_{\mu=1}^{M} h_\mu P_\mu, \qquad P_\mu\in\{I,X,Y,Z\}^{\otimes n}, \qquad H=H^\dagger .
    \label{eq:pauli_hamiltonian}
\end{equation}
In equation \eqref{eq:pauli_hamiltonian}, $H$ is the electronic qubit Hamiltonian, $M$ is the number of Pauli terms, $h_\mu\in\mathbb{R}$ is the coefficient of term $\mu$, and $P_\mu$ is an $n$-fold tensor product of one-qubit Pauli operators. In the numerical representation used for the benchmarks, $H$ is a dense Hermitian matrix in $\mathbb{C}^{D\times D}$ with Hilbert-space dimension $D=2^n$. When a scalar shift $s\in\mathbb{R}$ is present, it contains constant energy contributions such as nuclear repulsion and frozen-core terms.

For circuit parameters $\theta$, the variational state is denoted by $|\psi(\theta)\rangle$ and is normalized to unit length. The electronic expectation value minimized by the optimizer is
\begin{equation}
    e(\theta)=\langle\psi(\theta)|H|\psi(\theta)\rangle .
    \label{eq:electronic_energy}
\end{equation}
Here $e(\theta)$ is an electronic energy in Hartree (Ha), and the bra-ket notation denotes the standard complex inner product. The total molecular energy associated with the same state is
\begin{equation}
    E(\theta)=e(\theta)+s .
    \label{eq:total_energy}
\end{equation}
In equation \eqref{eq:total_energy}, $E(\theta)$ and $s$ are both measured in Ha. The scalar shift $s$ is not a variational parameter; it is a fixed scalar associated with the molecular Hamiltonian.

After completion of the prescribed optimization, exact diagonalization of $H$ is used to compute
\begin{equation}
    e_0=\lambda_{\min}(H), \qquad E_0=e_0+s .
    \label{eq:exact_energy}
\end{equation}
In equation \eqref{eq:exact_energy}, $e_0$ is the lowest electronic eigenvalue of $H$, $\lambda_{\min}$ denotes the minimum eigenvalue, and $E_0$ is the exact total energy under the same shift convention as equation \eqref{eq:total_energy}. The reported absolute energy error is
\begin{equation}
    \epsilon_{\Ha}=\left|E(\theta_\star)-E_0\right|=\left|e(\theta_\star)-e_0\right|, \qquad
    \epsilon_{\mHa}=1000\,\epsilon_{\Ha} .
    \label{eq:error_mha}
\end{equation}
In equation \eqref{eq:error_mha}, $\theta_\star$ is the final optimized parameter vector, $\epsilon_{\Ha}$ is the absolute error in Hartree (Ha), and $\epsilon_{\mHa}$ is the same quantity in milli-Hartree (mHa). The equality between the total-energy and electronic-energy errors holds because the same shift $s$ appears in both the variational and exact total energies. Chemical accuracy is assessed with the conventional approximate threshold $1.60\,\mHa=0.001600\,\Ha$ \cite{gonthier2022measurements,lim2024fragment}.

The state fidelity reported in the tables is
\begin{equation}
    F=\left|\langle\psi_0|\psi(\theta_\star)\rangle\right|^2 .
    \label{eq:fidelity}
\end{equation}
In equation \eqref{eq:fidelity}, $|\psi_0\rangle$ is the exact ground-state eigenvector returned by post-optimization diagonalization. For the non-degenerate benchmark cases considered here, $F$ measures how closely the optimized state matches the exact ground state up to an irrelevant global phase.

The optimization and evaluation phases use different information. During optimization, equation~\eqref{eq:electronic_energy} is the objective, and parameter updates use only variational energy values. The initialization chooses a computational basis state from diagonal Hamiltonian entries,
\begin{equation}
    b_0=\operatorname*{arg\,min}_{b\in\{0,1\}^{n}}\langle b|H|b\rangle .
    \label{eq:diag_init}
\end{equation}
In equation \eqref{eq:diag_init}, $|b\rangle$ is a computational basis vector indexed by a bit string $b$, and $b_0$ is the lowest-energy diagonal basis index. This rule uses only diagonal elements of the provided Hamiltonian matrix. The exact quantities $e_0$, $E_0$, $|\psi_0\rangle$, $\epsilon_{\Ha}$, $\epsilon_{\mHa}$, and $F$ are computed only after $\theta_\star$ has been fixed.

\section{Givens-exchange circuit construction}
\label{sec:construction}
The complex Stiefel manifold of $r$ orthonormal vectors in $D$ dimensions is
\begin{equation}
    \mathrm{St}_{\mathbb{C}}(D,r)=\{P\in\mathbb{C}^{D\times r}:P^\dagger P=I_r\} .
    \label{eq:stiefel}
\end{equation}
In equation \eqref{eq:stiefel}, $P$ is a matrix whose columns form an orthonormal frame, $P^\dagger$ is its conjugate transpose, and $I_r$ is the $r\times r$ identity. The experiments in this manuscript use $r=1$, so the frame is a single normalized state vector, but the Stiefel notation makes the unitary constraint explicit. If a unitary operation $U_t$ is applied at circuit step $t$, the frame evolves as
\begin{equation}
    P_{t+1}=U_tP_t, \qquad U_t^\dagger U_t=I_D .
    \label{eq:stiefel_update}
\end{equation}
In equation \eqref{eq:stiefel_update}, $P_t$ is the current frame, $P_{t+1}$ is the updated frame, $U_t$ is a $D\times D$ unitary matrix, and $I_D$ is the $D\times D$ identity. This update preserves $P_t^\dagger P_t=I_r$ and therefore keeps the variational state on the Stiefel manifold. The endpoint determines the energy, whereas the ordered path determines the implemented circuit structure. Stiefel and Grassmann geometry are standard tools for orthogonality-constrained eigenvalue problems \cite{edelman1998}.

The Stiefel viewpoint is used here as a coordinate-organizing principle rather than as a separate Riemannian optimization algorithm. For every computational-basis configuration of the spectator qubits, an exchange macro applies the same norm-preserving two-dimensional rotation to the corresponding $|\cdots 01\cdots\rangle$ and $|\cdots 10\cdots\rangle$ amplitudes. Globally, it is therefore a block-diagonal unitary on $\mathrm{St}_{\mathbb{C}}(D,1)$, rather than a single coordinate-plane rotation in the full $D$-dimensional state space. The ordered product of these blocks and the local rotations defines the structured circuit trajectory. The full circuit is not particle-number conserving, because the local $R_Y$ rotations can mix computational-basis sectors; the direct particle-conserving interpretation applies only to the Givens exchange macro in a \JW occupation basis.

The local rotation layer at circuit level $\ell$ is
\begin{equation}
    R_\ell=\prod_{q=0}^{n-1}R_Y^{(q)}(\alpha_{\ell q}), \qquad
    R_Y^{(q)}(\alpha)=\exp\left(-\frac{i\alpha}{2}Y_q\right) .
    \label{eq:ry_layer}
\end{equation}
In equation \eqref{eq:ry_layer}, $q$ indexes a qubit, $\alpha_{\ell q}$ is a real rotation angle, $Y_q$ is the Pauli $Y$ operator acting on qubit $q$ and as the identity on all other qubits, and $R_\ell$ is the product of all one-qubit $R_Y$ rotations in layer $\ell$. Rotations on different qubits commute, so their order inside $R_\ell$ does not affect the state.

For each pair of qubits $i<j$, the Givens exchange macro $G_{ij}(\beta)$ is defined on the two-qubit computational basis by
\begin{equation}
\begin{aligned}
G_{ij}(\beta)|00\rangle &= |00\rangle, &
G_{ij}(\beta)|11\rangle &= |11\rangle,\\
G_{ij}(\beta)|01\rangle
  &= \cos\beta\,|01\rangle-\sin\beta\,|10\rangle,\\
G_{ij}(\beta)|10\rangle
  &= \sin\beta\,|01\rangle+\cos\beta\,|10\rangle .
\end{aligned}
\label{eq:givens}
\end{equation}
In equation \eqref{eq:givens}, $\beta$ is a real exchange angle and the macro acts as the identity on all qubits except $i$ and $j$. The operation rotates the single-excitation subspace spanned by $|01\rangle$ and $|10\rangle$ while leaving $|00\rangle$ and $|11\rangle$ unchanged. Some software libraries parameterize the same two-dimensional exchange rotation with an angle differing by a sign or a factor of two; throughout this manuscript $\beta$ denotes the matrix angle shown explicitly in equation \eqref{eq:givens}. In a \JW occupation encoding this is aligned with number-preserving single-excitation structure, whereas in parity or tapered encodings it should be interpreted as a valid variational macro rather than as a literal fermionic excitation.

An exchange block at level $\ell$ is
\begin{equation}
\begin{aligned}
G_\ell
  &= \prod_{(i,j)\in\mathcal{E}_{\mathrm{lex}}}
     G_{ij}(\beta_{\ell ij}),\\
\mathcal{E}_{\mathrm{lex}}
  &= \{(0,1),(0,2),\ldots,(n-2,n-1)\} .
\end{aligned}
\label{eq:givens_block}
\end{equation}
In equation \eqref{eq:givens_block}, $\mathcal{E}_{\mathrm{lex}}$ is the fixed lexicographic all-pair order defining the exchange block, and $\beta_{\ell ij}$ is the angle for pair $(i,j)$ in block $\ell$. Two Givens macros sharing a qubit generally do not commute, so the ordered set $\mathcal{E}_{\mathrm{lex}}$ is part of the ansatz definition.

With $L$ exchange blocks, the implemented state is
\begin{equation}
    |\psi(\theta)\rangle=R_LG_{L-1}R_{L-1}\cdots G_1R_1G_0R_0|b_0\rangle .
    \label{eq:implemented_ansatz}
\end{equation}
In equation \eqref{eq:implemented_ansatz}, $|b_0\rangle$ is the diagonal initialization from equation \eqref{eq:diag_init}, and $\theta$ collects all rotation angles $\alpha_{\ell q}$ and exchange angles $\beta_{\ell ij}$. The experiments use $L=2$, giving two all-pair exchange blocks and three local $R_Y$ layers. The maximum number of exchange macros is $L n(n-1)/2$, and the maximum number of $R_Y$ rotations is $(L+1)n$. All angles are initialized from a zero-mean normal distribution with scale $0.05$ and then updated with Adam. A retained macro count is computed only after optimization by applying the activity threshold
\begin{equation}
\begin{aligned}
N_{\mathrm{Givens}}
  &= \#\{(\ell,i,j):|\beta_{\ell ij}|>\tau\},\\
N_{\mathrm{ROT}}
  &= \#\{(\ell,q):|\alpha_{\ell q}|>\tau\},\\
\tau &= 10^{-4} .
\end{aligned}
\label{eq:resource_counts}
\end{equation}
In equation \eqref{eq:resource_counts}, $N_{\mathrm{Givens}}$ is the reported number of active Givens exchange macros, $N_{\mathrm{ROT}}$ is the reported number of active one-qubit rotations, $\#$ denotes set cardinality, and $\tau$ is the fixed activity threshold used for resource reporting. These are variational macro counts, not hardware-native gate counts.

\section{Benchmark protocol}
The primary benchmark comprises three molecular labels under the stored Hamiltonian conventions: LiH-6 at $2.2$~\AA{} with \JW mapping and no tapering, \BeHtwo-6 with \JW mapping and no tapering, and \HtwoO-8 with \JW mapping and no tapering. These labels permit the most direct evaluation of the present ansatz. For LiH-6 at $2.2$~\AA{} and \HtwoO-8, the dense Hamiltonian matrices and scalar shifts were verified against the reference benchmark coefficient files; the maximum matrix difference, Frobenius-norm difference, shift difference and total exact-energy difference were all zero under the stored dense-matrix convention. The LiH-4 parity Hamiltonian at $3.4$~\AA{} was also verified against the corresponding reference coefficient file, but it is retained as a separate stress case rather than as the main LiH comparison. This separation is important because LiH-4 and LiH-6 differ not only in qubit number, but also in geometry, mapping, and tapering. A four-qubit parity-tapered Hamiltonian is not a smaller version of the six-qubit \JW Hamiltonian; it is a different encoded variational landscape with different symmetry reduction and operator structure. The main LiH comparison is therefore the LiH-6 \JW row, while the LiH-4 row is used to probe robustness under a harder parity-mapped and tapered setting.

The LiH and \HtwoO Hamiltonians are evaluated as dense Hermitian matrices under their stated mapping and shift conventions. The reference benchmark Hamiltonians are the dense LiH-4, LiH-6 and \HtwoO-8 operators used in the HyRLQAS quantum-architecture-search comparison \cite{hyrlqas}; the coefficient archive and resulting matrix-level identity reports are available through the public project repository identified in the Data Availability statement. The coefficient-identity check reported in Supplementary Table S6 supports direct same-Hamiltonian interpretation for the LiH-4 stress case and for the LiH-6 and \HtwoO-8 reference comparisons under the supplied dense-matrix convention. The \BeHtwo Hamiltonian is a public-specification candidate generated from the published molecular specification: neutral singlet \BeHtwo in the STO-3G basis with Be at $(0,0,0)$ and H atoms at $(0,0,\pm1.33)$~\AA{}, \JW mapping, and no tapering. The candidate uses four active electrons in three spatial orbitals, which yields six spin orbitals and therefore six qubits under \JW mapping. This label remains explicit in all tables because the public specification fixes the molecule, basis, mapping, and tapering settings, while the generated dense Hamiltonian is a reproducible reconstruction of that specification rather than a verified coefficient-identical reference file.

All reported Givens-exchange runs use the same numerical protocol: $L=2$ exchange blocks, local $R_Y$ rotations only, 150 Adam steps, learning rate $0.05$, one restart, diagonal-basis initialization, and the six seeds $17,11111,22222,33333,44444,55555$. Each seed is summarized by 161 reported energy evaluations: 150 gradient-bearing optimization evaluations and 11 fixed post-update monitoring evaluations at steps $0,15,30,45,60,75,90,105,120,135$ and $149$. The term ``one restart'' means that one independently initialized optimization trajectory is run for each seed. The main tables report six-seed mean errors, sample standard deviations and medians for this method, rather than selecting a different best seed for each molecule. Best and worst seeds remain in the sensitivity table to document the optimization range. This convention is conservative for point comparison with architecture-search studies that usually report selected optimized outcomes, but it gives a more direct estimate of the method's typical performance under the fixed seed set.

Resource columns are interpreted at the level at which they are measured. Givens counts are pre-compilation two-qubit exchange macro counts from equations \eqref{eq:givens} and \eqref{eq:resource_counts}. ROT denotes retained single-qubit rotations after applying the activity threshold. The published quantum-architecture-search baselines report CNOT and rotation counts in their own circuit representations. The resource figure therefore separates reported structural operation counts by reporting level rather than placing them on a single hardware-equivalent axis. The energy conclusions are independent of any hardware-native interpretation of a Givens macro.

The dedicated multi-method comparisons use the two primary coefficient-verified molecular labels, LiH-6 and \HtwoO-8. Published point errors for HyRLQAS, TensorRL-QAS, CRLQAS, Vanilla reinforcement learning, DQN-rank, TPPO and quantumDARTS are expressed in mHa after applying $\epsilon_{\mHa}=1000\epsilon_{\Ha}$, the same conversion used in Eq.~\eqref{eq:error_mha} \cite{hyrlqas,tensorrl,crlqas,vanillarl,benchrlqas,quantumdarts}. The LiH-6 and \HtwoO-8 Hamiltonians used here pass the matrix-level identity check against the reference benchmark Hamiltonians used in the HyRLQAS comparison; LiH-4 is included only as a coefficient-verified parity/taper stress context, and \BeHtwo is included only as a public-specification candidate because a coefficient-identical reference file was not supplied.

\section{Results}
Table~\ref{tab:primary} reports the primary six-seed average results for this method. The rows are LiH-6, the \BeHtwo public-specification candidate, and \HtwoO-8. The columns list the number of qubits $q$, the mapping and tapering convention, the exact total energy $E_0$ from Eq.~\eqref{eq:exact_energy}, the mean and standard deviation of the absolute error in mHa, the median mHa error, the pass count against chemical accuracy, the mean fidelity $F$ from Eq.~\eqref{eq:fidelity}, and retained mean macro counts. The exact conversion to Ha is given in Eq.~\eqref{eq:error_mha}; Table~\ref{tab:sensitivity} gives the associated best--worst seed range, and Tables~\ref{tab:lih6comparison} and \ref{tab:h2ocomparison} give the dedicated coefficient-verified LiH-6 and \HtwoO-8 comparisons using the same six-seed mean convention for this work.

\begin{table*}[!t]
\caption{Primary six-seed average Givens-exchange results. $E_0$ is the exact total energy after applying the stored shift. Errors are absolute and reported in mHa; standard deviations are sample standard deviations over the six fixed seeds. Givens counts are pre-compilation exchange macros, and ROT is the retained one-qubit rotation count averaged across seeds. The BeH$_2$ row is a public-specification candidate.}
\label{tab:primary}
\centering
\footnotesize
\setlength{\tabcolsep}{3.2pt}
\renewcommand{\arraystretch}{1.08}
\begin{tabularx}{\textwidth}{@{}>{\raggedright\arraybackslash}X c c c c c c c c@{}}
\toprule
Case & $q$ & map/taper & $E_0$ (Ha) & \shortstack{mean $\pm$ std\\error (mHa)} & \shortstack{median\\(mHa)} & pass & mean $F$ & \shortstack{mean\\Givens/ROT} \\
\midrule
LiH-6 JW $2.2$ & 6 & JW/no & $-7.844879127$ & $0.000124\pm0.000113$ & $0.000091$ & 6/6 & $1.000000$ & 30.00/17.33 \\
BeH$_2$-6 candidate & 6 & JW/no & $-15.563137014$ & $0.002152\pm0.004242$ & $0.000527$ & 6/6 & $0.999997$ & 30.00/17.50 \\
H$_2$O-8 JW & 8 & JW/no & $-73.294106759$ & $0.128558\pm0.133018$ & $0.090323$ & 6/6 & $0.999953$ & 56.00/24.00 \\
\bottomrule
\end{tabularx}
\end{table*}

The primary table shows that the six-seed means are below the $1.60\,\mHa$ chemical-accuracy threshold for all three rows, and every individual seed passes for these rows. Values in Ha can be recovered exactly from $\epsilon_{\Ha}=\epsilon_{\mHa}/1000$, so a redundant Ha column is omitted. This table is intentionally limited to the Hamiltonians evaluated in the present study. Direct external comparison is separated into the LiH-6 and \HtwoO-8 tables below, while the \BeHtwo row remains labelled as a public-specification candidate because coefficient identity with the HyRLQAS internal Hamiltonian is not claimed.

Table~\ref{tab:sensitivity} reports only the present method across the six seeds. For each row, the mean, sample standard deviation, and median summarize the distribution of $\epsilon_{\mHa}$, the best and worst entries give both error and seed number, the pass column gives the number of seeds below chemical accuracy, and the final two columns report mean fidelity and mean wall time.

\begin{table*}[!t]
\caption{Six-seed sensitivity analysis for the Givens-exchange method. Errors are absolute values in mHa. The best and worst columns report the error followed by the seed. The pass column gives the number of seeds below $1.60\,\mHa$.}
\label{tab:sensitivity}
\centering
\footnotesize
\setlength{\tabcolsep}{3.2pt}
\renewcommand{\arraystretch}{1.08}
\begin{tabularx}{\textwidth}{@{}>{\raggedright\arraybackslash}X c c c c c c c@{}}
\toprule
Case & \shortstack{mean $\pm$ std\\error} & median & best (seed) & worst (seed) & pass & mean $F$ & \shortstack{mean wall\\(s)} \\
\midrule
LiH-6 JW $2.2$ & $0.000124\pm0.000113$ & $0.000091$ & $0.000015$ (22222) & $0.000292$ (55555) & 6/6 & $1.000000$ & 2.83 \\
BeH$_2$-6 candidate & $0.002152\pm0.004242$ & $0.000527$ & $0.000132$ (55555) & $0.010802$ (17) & 6/6 & $0.999997$ & 3.66 \\
H$_2$O-8 JW & $0.128558\pm0.133018$ & $0.090323$ & $0.022656$ (44444) & $0.377337$ (22222) & 6/6 & $0.999953$ & 4.94 \\
LiH-4 parity stress & $0.555932\pm0.898445$ & $0.146842$ & $0.000158$ (44444) & $2.294095$ (17) & 5/6 & $0.945242$ & 1.49 \\
\bottomrule
\end{tabularx}
\end{table*}

The sensitivity analysis underlies the point comparison reported above. LiH-6, the \BeHtwo candidate, and \HtwoO-8 pass chemical accuracy in every seed. Their mean fidelities are close to one, indicating that the low energy errors correspond to trial states close to the exact ground states of the stored Hamiltonians. The LiH-4 parity stress row is qualitatively different: it has chemically accurate seeds, including a very low best error, but also one high-error local solution. This behaviour supports treating LiH-4 as a stress test for optimization robustness rather than as the primary LiH comparison.

Table~\ref{tab:stressscope} reports the main scope tests outside the primary comparison. These rows are not used as external comparison claims. They show how the fixed exchange prior behaves when the mapping or molecular regime changes.

\begin{table*}[!t]
\caption{Scope tests under stretched geometry and parity mapping. All rows use the same optimization protocol as the primary cases but are interpreted as stress tests rather than external-comparison claims. Errors are absolute values in mHa.}
\label{tab:stressscope}
\centering
\footnotesize
\setlength{\tabcolsep}{4pt}
\renewcommand{\arraystretch}{1.08}
\begin{tabularx}{\textwidth}{@{}>{\raggedright\arraybackslash}X c c c c c c c@{}}
\toprule
Case & $q$ & map/taper & best (seed) & mean $\pm$ std & median & pass & mean $F$ \\
\midrule
LiH-4 parity $3.4$~\AA{} & 4 & parity/yes & $0.000158$ (44444) & $0.555932\pm0.898445$ & $0.146842$ & 5/6 & $0.945242$ \\
LiH-6 JW $3.4$~\AA{} & 6 & JW/no & $0.006796$ (17) & $6.912731\pm3.385779$ & $8.353566$ & 1/6 & $0.176254$ \\
LiH-6 parity $3.4$~\AA{} & 6 & parity/yes & $8.369536$ (17) & $8.371899\pm0.002826$ & $8.370691$ & 0/6 & $0.000000$ \\
\bottomrule
\end{tabularx}
\end{table*}

The stress tests delineate the scope of the method. The ansatz can reach a chemically accurate LiH-4 parity solution in several seeds, but it is sensitive to initialization. At the stretched LiH-6 geometry, the \JW row passes in only one of six seeds and the parity row does not pass. This behaviour is consistent with the intended scope of the exchange prior: it is strongest when the qubit representation retains a direct occupation-like interpretation and the electronic state is not dominated by a harder stretched-bond correlation pattern.

A compact depth ablation in Supplementary Table S3 compares $L=0$, $L=1$, and $L=2$ under the same six seeds and optimizer settings. The $L=0$ setting is a diagonal-initialized local-rotation control with no exchange macros; it remains above chemical accuracy for all four analysed labels. One exchange block improves LiH-4 and LiH-6 but does not make \BeHtwo or \HtwoO chemically accurate. For \BeHtwo, the $L=1$ row remains essentially at the $L=0$ error scale, indicating that the optimizer remains close to the diagonal-initialized basin until the second exchange block is added. Two exchange blocks are therefore the smallest depth tested here that gives chemical accuracy for all six seeds of LiH-6, \BeHtwo, and \HtwoO and for five of six LiH-4 stress seeds. This ablation supports the interpretation that the exchange layer, not only diagonal initialization or local rotations, is responsible for the reported molecular accuracy.

Tables~\ref{tab:lih6comparison} and \ref{tab:h2ocomparison} provide dedicated coefficient-verified comparisons against published quantum architecture search methods for LiH-6 and \HtwoO-8. Each table reports absolute error in mHa; values taken from Hartree-scale source tables are converted using Eq.~\eqref{eq:error_mha}. The Givens-exchange entries use the six-seed mean error and mean retained rotation count. Their depths are listed at the macro level because the present resource accounting is at the Givens-exchange level; the other rows use the depth, CNOT and rotation values reported for the corresponding published methods.

\begin{table*}[!t]
\caption{Coefficient-verified LiH-6 comparison with published QAS methods. The dense LiH-6 Hamiltonian and scalar shift used here are identical to the reference benchmark used in the HyRLQAS comparison \cite{hyrlqas} under the stored dense-matrix convention. Errors are absolute values in mHa. The Givens entry is a pre-compilation macro count; comparison rows report CNOT counts.}
\label{tab:lih6comparison}
\centering
\footnotesize
\setlength{\tabcolsep}{3.6pt}
\renewcommand{\arraystretch}{1.08}
\begin{tabularx}{\textwidth}{@{}>{\raggedright\arraybackslash}p{0.27\textwidth}>{\raggedright\arraybackslash}X c c c c@{}}
\toprule
Method & class & error (mHa) & depth & two-qubit report & ROT \\
\midrule
Givens exchange, six-seed mean & fixed exchange ansatz & $0.000124$ & macro & 30 Givens & 17.33 \\
HyRLQAS \cite{hyrlqas} & hybrid-action RL-QAS & $0.590000$ & 23 & 27 CNOT & 24 \\
TensorRL(HyRLQAS) \cite{tensorrl,hyrlqas} & TensorRL-QAS & $0.810000$ & 5 & 5 CNOT & 4 \\
TensorRL(CRLQAS) \cite{tensorrl,crlqas} & TensorRL-QAS & $1.300000$ & 6 & 6 CNOT & 4 \\
CRLQAS \cite{crlqas} & curriculum RL-QAS & $1.500000$ & 29 & 28 CNOT & 22 \\
Vanilla RL \cite{vanillarl} & RL-QAS & $3.700000$ & 36 & 38 CNOT & 25 \\
DQN-rank \cite{benchrlqas} & BenchRL-QAS & $1.200000$ & 24 & 19 CNOT & 43 \\
TPPO \cite{benchrlqas} & BenchRL-QAS & $1.000000$ & 25 & 23 CNOT & 42 \\
quantumDARTS \cite{quantumdarts} & differentiable QAS & $0.290000$ & 54 & 52 CNOT & 80 \\
\bottomrule
\end{tabularx}
\end{table*}

The coefficient-verified LiH-6 comparison shows that the six-seed mean remains several orders of magnitude below the chemical-accuracy threshold and lower than the LiH-6 point errors listed for HyRLQAS, TensorRL-QAS, CRLQAS, Vanilla reinforcement learning, DQN-rank, TPPO and quantumDARTS. The reported six-seed mean is $0.000124\,\mHa$. The table also shows the complementary resource profile: several search methods report fewer CNOTs, while the present method uses a systematic all-pair Givens exchange pattern. The result is therefore best interpreted as an accuracy-focused point in the ansatz-design trade-off, not as a hardware-resource advantage.

\begin{table*}[!t]
\caption{Coefficient-verified H$_2$O-8 comparison with published QAS methods. The dense H$_2$O-8 Hamiltonian and scalar shift used here are identical to the reference benchmark used in the HyRLQAS comparison \cite{hyrlqas} under the stored dense-matrix convention. Errors are absolute values in mHa. The Givens error is the six-seed mean; the best seed is $0.022656\,\mHa$ at seed 44444. Givens macros and CNOTs are not hardware-equivalent.}
\label{tab:h2ocomparison}
\centering
\footnotesize
\setlength{\tabcolsep}{3.6pt}
\renewcommand{\arraystretch}{1.08}
\begin{tabularx}{\textwidth}{@{}>{\raggedright\arraybackslash}p{0.27\textwidth}>{\raggedright\arraybackslash}X c c c c@{}}
\toprule
Method & class & error (mHa) & depth & two-qubit report & ROT \\
\midrule
Givens exchange, six-seed mean & fixed exchange ansatz & $0.128558$ & macro & 56 Givens & 24.00 \\
HyRLQAS \cite{hyrlqas} & hybrid-action RL-QAS & $0.170000$ & 23 & 25 CNOT & 14 \\
TensorRL(HyRLQAS) \cite{tensorrl,hyrlqas} & TensorRL-QAS & $0.510000$ & 4 & 6 CNOT & 4 \\
TensorRL(CRLQAS) \cite{tensorrl,crlqas} & TensorRL-QAS & $0.910000$ & 11 & 8 CNOT & 10 \\
CRLQAS \cite{crlqas} & curriculum RL-QAS & $1.100000$ & 30 & 25 CNOT & 42 \\
Vanilla RL \cite{vanillarl} & RL-QAS & $1.100000$ & 59 & 67 CNOT & 65 \\
DQN-rank \cite{benchrlqas} & BenchRL-QAS & $0.920000$ & 35 & 34 CNOT & 30 \\
TPPO \cite{benchrlqas} & BenchRL-QAS & $0.950000$ & 56 & 76 CNOT & 29 \\
quantumDARTS \cite{quantumdarts} & differentiable QAS & $0.310000$ & 64 & 68 CNOT & 151 \\
\bottomrule
\end{tabularx}
\end{table*}

The H$_2$O-8 comparison gives the same qualitative pattern on a second coefficient-verified Hamiltonian. The Givens-exchange six-seed mean, $0.128558\,\mHa$, is lower than the published point errors in the comparison set, while the all-pair template uses more two-qubit macros than the compact searched circuits report as CNOT counts. The H$_2$O-8 margin is modest relative to seed variation, so the conclusion is strongest when read together with the LiH-6 comparison and the full six-seed sensitivity table.

Table~\ref{tab:contextfour} reports the four benchmark labels in mHa under a common unit convention. This table is contextual: LiH-6 and H$_2$O-8 are coefficient-verified primary comparisons, LiH-4 is a coefficient-verified parity/tapered stress case, and \BeHtwo is a public-specification candidate without coefficient-identity verification.

\begin{table*}[!t]
\caption{Contextual four-label comparison in mHa. Present-method values are six-seed means; HyRLQAS values are published point errors. LiH-6 and H$_2$O-8 support coefficient-identical primary comparisons, LiH-4 is a coefficient-verified stress case with different geometry and encoding, and BeH$_2$ is a public-specification candidate.}
\label{tab:contextfour}
\centering
\footnotesize
\setlength{\tabcolsep}{3.5pt}
\renewcommand{\arraystretch}{1.12}
\begin{tabularx}{\textwidth}{@{}>{\raggedright\arraybackslash}p{0.15\textwidth}>{\raggedright\arraybackslash}p{0.22\textwidth} c c >{\raggedright\arraybackslash}X@{}}
\toprule
Case & comparison status & \shortstack{this work\\mean (mHa)} & \shortstack{HyRLQAS\\(mHa)} & interpretation \\
\midrule
LiH-4 parity $3.4$~\AA{} & coefficient-verified stress case & $0.555932$ & $0.000012$ & HyRLQAS is lower; this row documents seed-sensitive parity/taper performance and is not the primary LiH claim. \\
LiH-6 JW $2.2$~\AA{} & coefficient-verified primary comparison & $0.000124$ & $0.590000$ & This work has the lower six-seed mean under the shared Hamiltonian convention. \\
BeH$_2$-6 & public-specification candidate & $0.002152$ & $0.000063$ & Label-level context only; coefficient identity is not claimed for this row. \\
H$_2$O-8 JW & coefficient-verified primary comparison & $0.128558$ & $0.170000$ & This work has the lower six-seed mean, with a modest margin relative to seed variation. \\
\bottomrule
\end{tabularx}
\end{table*}

\begin{figure*}[!t]
\centering
\includegraphics[width=0.93\textwidth]{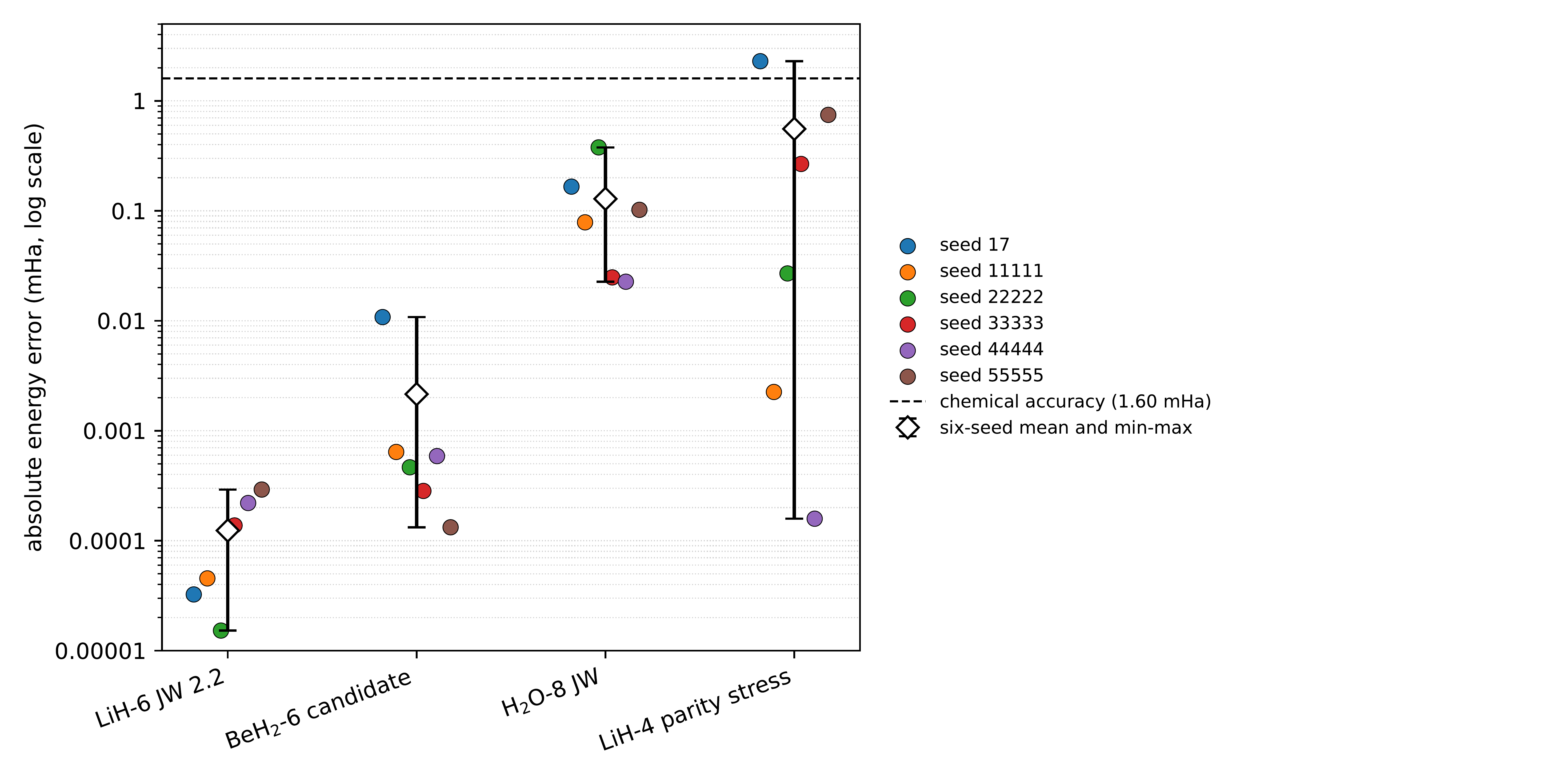}
\caption{Six-seed energy sensitivity of the Givens-exchange method. The y-axis reports absolute energy error in mHa with logarithmic spacing and decimal tick labels. Coloured points show individual seeds, open diamonds show six-seed means, vertical intervals show min--max ranges, and the dashed line marks chemical accuracy at $1.60\,\mHa$. LiH-4 is labelled as a parity-mapped stress case rather than as the primary LiH comparison.}
\label{fig:stiefel_distribution}
\end{figure*}

Fig.~\ref{fig:stiefel_distribution} visualizes the seed distribution for this method only. The individual points are the six seeds, the open diamond is the six-seed mean, the vertical line spans the minimum to maximum error, and the dashed line marks chemical accuracy. The axis uses logarithmic spacing because the values span several orders of magnitude, while decimal tick labels preserve a common numerical format. The legend is placed outside the plotting region.

The figure shows that the primary rows are robustly below chemical accuracy across all seeds. It also shows why reporting the LiH-4 distribution is more informative than reporting a single LiH-4 point: the distribution spans chemically accurate and non-chemically accurate outcomes, which is characteristic of a local-optimization sensitivity rather than a simple absence of expressivity.

\begin{figure*}[!t]
\centering
\includegraphics[width=0.93\textwidth]{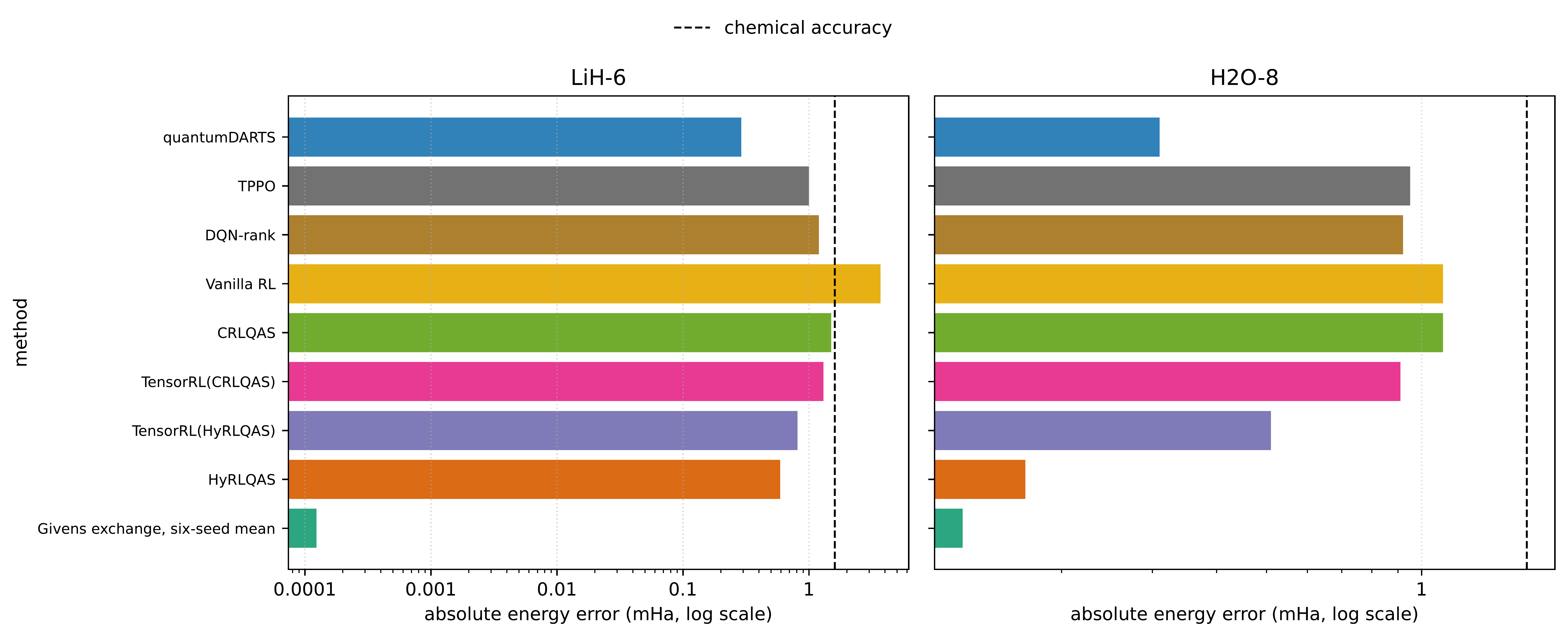}
\caption{Energy-error comparison on the two coefficient-verified Hamiltonians. The Givens-exchange values are six-seed means. Published comparison values are point errors in mHa, matching the conventions in Tables~\ref{tab:lih6comparison} and \ref{tab:h2ocomparison}. The x-axes use logarithmic spacing with decimal tick labels, and the dashed line marks chemical accuracy at $1.60\,\mHa$.}
\label{fig:verifiedenergy}
\end{figure*}

Fig.~\ref{fig:verifiedenergy} gives a second view of the LiH-6 and H$_2$O-8 comparisons by plotting energy error for the methods in Tables~\ref{tab:lih6comparison} and \ref{tab:h2ocomparison}. Logarithmic spacing is used because the errors span several orders of magnitude, and decimal tick labels match the table convention.

\begin{figure*}[!t]
\centering
\includegraphics[width=0.93\textwidth]{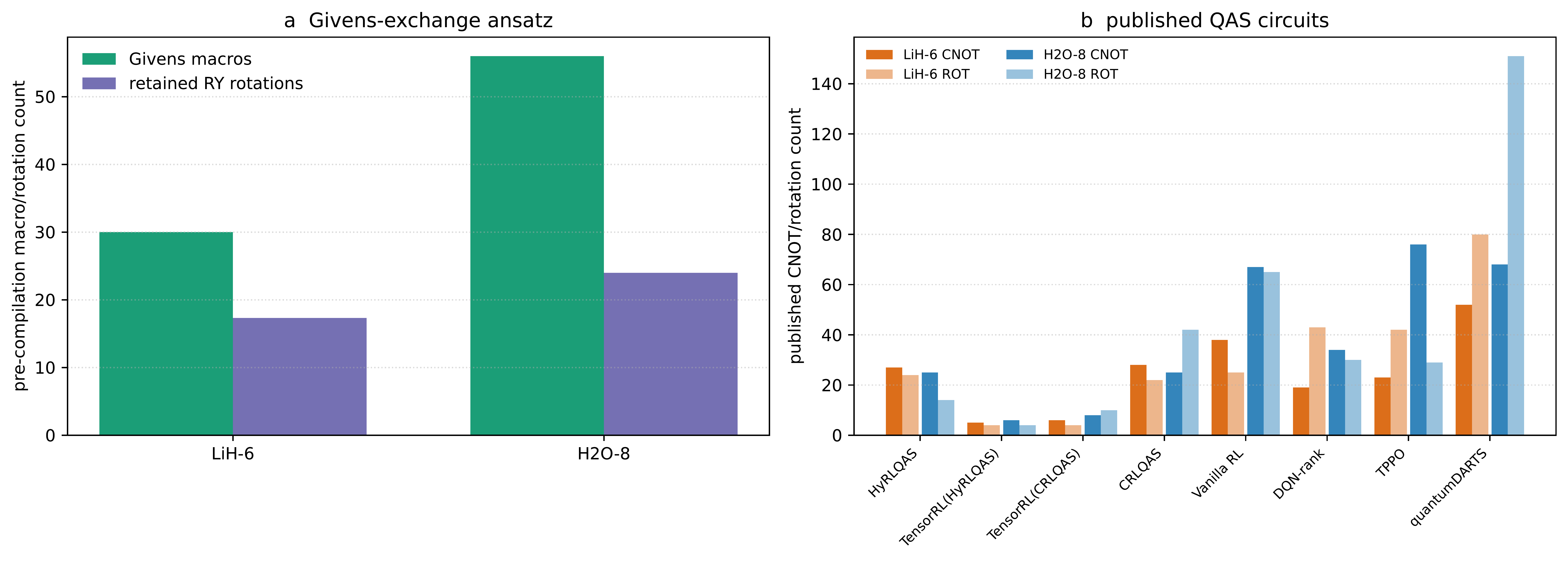}
\caption{Resource counts separated by reporting level for the two coefficient-verified comparisons. Panel a reports this method's pre-compilation Givens exchange macro count and retained one-qubit rotations. Panel b reports the CNOT and rotation counts of the published comparison methods for LiH-6 and H$_2$O-8. The two panels should not be interpreted as a hardware-equivalent compiled-depth comparison.}
\label{fig:verifiedresources}
\end{figure*}

Fig.~\ref{fig:verifiedresources} focuses the resource visualization on the same two verified comparisons while separating reporting levels. Panel a shows this method's pre-compilation Givens macro and rotation counts. Panel b shows the CNOT and rotation counts reported for the published comparison methods. The panels are not a native-gate equivalence claim; they identify the level at which each resource number is defined.

The resource figure reinforces the interpretation of Tables~\ref{tab:lih6comparison} and \ref{tab:h2ocomparison}. The present method obtains lower mean energy errors on the two coefficient-verified comparisons, while several architecture-search baselines report shallower or fewer-CNOT circuits in their native reporting. The separated panels avoid equating a Givens macro with a CNOT and motivate native-gate compilation as the next resource-focused refinement.

\section{Discussion}
The central result is that a fixed Givens-exchange topology can reach highly accurate molecular energies without learning a gate-placement policy. In the primary rows, LiH-6, \HtwoO-8 and the \BeHtwo public-specification candidate pass chemical accuracy in every seed, and their six-seed means are substantially below the chemical-accuracy threshold. The dedicated LiH-6 and \HtwoO-8 tables place this result against recent quantum architecture search methods and show that the Givens-exchange means are lower than the published point errors on both coefficient-verified primary comparisons. The contextual four-label table deliberately also shows LiH-4 and \BeHtwo: LiH-4 is included to document the stress-case limitation, and \BeHtwo is included only as public-specification context rather than as a coefficient-identical comparison.

The comparison should be read as an energy-resource trade-off. The present ansatz uses an all-pair exchange template and therefore has a larger macro-operation budget than some learned compact circuits. This is the price paid for avoiding architecture search and for using a transparent, systematically defined exchange prior. Consequently, the main claim is not that the ansatz is a hardware-resource winner. Rather, the result shows that a simple fixed molecular template can provide a highly accurate reference point against which search-based methods and future compiled variants can be evaluated.

A complementary efficiency axis is the absence of an architecture-search loop. Reinforcement-learning quantum architecture search constructs circuits through repeated decision steps and then uses an external classical optimizer to refine candidate circuit parameters before retaining a policy or circuit for final evaluation \cite{hyrlqas,crlqas,benchrlqas}. The present ansatz fixes the topology before optimization, so each reported seed requires one variational optimization rather than a learned circuit-construction process. In the current noiseless dense-state simulations, the six-seed mean wall times are $2.83$~s for LiH-6 and $4.94$~s for H$_2$O-8. These wall times should not be interpreted as hardware-independent speedups, because the compared methods were developed and profiled under different software and hardware conditions. They do show, however, that the accuracy reported here is obtained without the additional policy-training or architecture-search cost that search-based methods require.

The present numerical evidence is limited to noiseless dense-state simulation and therefore does not establish hardware-level energy accuracy. On a quantum processor, finite-shot estimation, decoherence, two-qubit-gate errors, connectivity constraints, and routing of the all-pair graph can alter both the optimization landscape and the final energy. For $L=2$, the pre-compilation ansatz contains $n(n-1)$ exchange parameters and $3n$ local-rotation parameters, for a total of $n^2+2n$ parameters: 48 at six qubits and 80 at eight qubits. The topology is fixed across seeds, and Table~\ref{tab:primary} reports retained counts close to the corresponding maxima (30 Givens macros and 18 $R_Y$ rotations at six qubits; 56 Givens macros and 24 $R_Y$ rotations at eight qubits). The available results therefore do not show that the lowest-error seed uses a structurally simpler circuit; a per-seed relationship between error and compiled native-gate complexity was not measured. Hardware assessment requires mapping-aware native compilation and routing, followed by finite-shot or noisy reoptimization and direct comparison after compilation. Symmetry checking, noise-aware parameter prediction, and low-depth virtual distillation provide relevant validation or mitigation tools for such a study \cite{zhang2025symmetry,karim2025mlvqe,karim2024virtual}, while noisy-processor electronic-structure calculations illustrate the importance of evaluating the complete workflow on hardware \cite{zhang2024band}.

The scope tests are central to the interpretation. LiH-4 parity at $3.4$~\AA{} and LiH-6 \JW at $2.2$~\AA{} are not interchangeable LiH benchmarks because they differ in geometry, mapping, tapering, qubit count, and operator structure, even when the LiH-4 Hamiltonian itself is coefficient-verified against its reference file. The broader stress table shows that the method is strongest for near-equilibrium \JW active-space Hamiltonians and less robust for stretched LiH and parity-mapped effective Hamiltonians. This behaviour is consistent with the construction: the Givens exchange macro has a direct single-excitation interpretation in a \JW occupation basis, whereas after parity mapping or tapering it remains a valid variational operation but no longer represents the same literal orbital exchange.

The depth ablation in the supplementary information addresses the mechanism behind the result. The diagonal starting state and local rotations alone do not reproduce the two-layer performance. One exchange block improves some rows but is not uniformly chemically accurate, and the \BeHtwo one-layer row remains close to the diagonal-initialized basin. Two ordered all-pair exchange blocks provide the expressivity that produces the reported primary accuracy. This supports the geometric interpretation that molecular accuracy arises from structured exchange-induced amplitude redistribution on the state manifold rather than from diagonal initialization alone.

The Stiefel language is deliberately limited. The calculation does not perform Riemannian optimization on the Stiefel manifold. Instead, the Stiefel viewpoint explains why the ordered exchange sequence can be regarded as a product of low-dimensional coordinate-plane rotations acting on a normalized state. This interpretation helps distinguish the ansatz from an unconstrained hardware-efficient layer, but the numerical optimization itself is the ordinary variational optimization of the circuit angles.

The resource accounting is deliberately conservative. Givens macros are useful at the ansatz-design level because they define a transparent exchange operation, whereas hardware execution would require compilation to a native gate basis. The current resource comparison therefore reports macro counts for this work and CNOT counts for the published comparison methods at their respective reporting levels. This framing preserves the energy result while identifying native-gate compilation and reoptimization as the appropriate next resource-focused study.

The \BeHtwo row is also framed at the scientifically appropriate level. It is a public-specification candidate generated from the molecule, basis, mapping, and tapering information available for the benchmark label. The exact energy and errors are exact for the Hamiltonian used here, and the Hamiltonian construction is reproducible. The candidate label reflects the fact that public molecular settings do not by themselves prove coefficient identity with an independently generated Pauli coefficient file. By contrast, the LiH-4 stress, LiH-6 and \HtwoO-8 dense Hamiltonians and shifts were checked against reference benchmark coefficient files and matched at matrix level, as reported in the supplementary information.

Taken together, the results support the use of a Givens-exchange ansatz with a Stiefel-coordinate interpretation as a transparent molecular prior. Its main near-term development path is to keep the same exchange logic while reducing the all-pair graph, compiling the retained macros to a native gate basis, and reoptimizing after compilation. One natural pair-selection score for a Pauli Hamiltonian $H=\sum_\mu h_\mu P_\mu$ is
\begin{equation}
    A_{ij}=\sum_{\mu:\,P_{\mu i}\neq I,\,P_{\mu j}\neq I}|h_\mu|,
    \label{eq:pair_score}
\end{equation}
where $P_{\mu i}$ and $P_{\mu j}$ are the single-qubit Pauli factors of string $P_\mu$ on qubits $i$ and $j$. Equation~\eqref{eq:pair_score} was not used to select pairs in the reported all-pair runs; it gives a reproducible future criterion for prioritizing qubit pairs that co-occur strongly in the molecular Hamiltonian. Because this score counts all Pauli co-occurrences, including diagonal terms, a more chemistry-specific future variant could restrict the sum to hopping-like $XX$, $YY$ or mixed excitation-relevant terms.

\section{Conclusion}
This work presents a fixed-topology Givens-exchange ansatz for molecular variational eigensolvers and evaluates it under an explicit separation of variational optimization from post-optimization benchmarking. The method prepares a diagonal computational-basis state, applies local $R_Y$ layers and ordered Givens exchange blocks, and optimizes Hamiltonian expectation values with a two-layer protocol. The primary six-seed average results show chemically accurate LiH-6, \HtwoO-8 and \BeHtwo public-specification candidate energies, and all three labels pass chemical accuracy in six of six seeds. Matrix-level checks verify coefficient identity for the LiH-6 and \HtwoO-8 Hamiltonians against the reference benchmark Hamiltonians used in the HyRLQAS comparison, while \BeHtwo remains appropriately labelled as a public-specification candidate. LiH-4 is included as a coefficient-verified parity/tapered stress case and not as the primary LiH comparison. In the two primary coefficient-verified comparisons, the Givens-exchange six-seed means are lower than the published point errors of the compared quantum architecture search methods when expressed consistently in mHa. The method is therefore best viewed as an accurate, interpretable and search-free ansatz-design template. Its development path is to retain the exchange logic while replacing the all-pair graph with a mapping-aware Hamiltonian-informed subset, compiling retained macros to a native gate basis, and reoptimizing after compilation.

\section*{Data Availability}

The molecular Hamiltonians, scalar shifts, mapping and tapering metadata, final optimized parameter sets, seed-level numerical results, Hamiltonian-identity reports, and data underlying all tables and figures are available in the \href{https://github.com/PR2aid/A-Givens-exchange-ansatz-for-molecular-variational-eigensolvers/tree/main}{public project repository}. These artifacts are sufficient to reproduce every table, figure, and Hamiltonian-identity check reported in the manuscript and supplementary information. No proprietary, private, or restricted dataset is required to evaluate the claims reported in this manuscript.

\section*{Code Availability}

A standalone utility for the Hamiltonian-identity checks and the scripts that regenerate the reported tables and figures from the deposited numerical results are included in the following archive \href{https://github.com/PR2aid/A-Givens-exchange-ansatz-for-molecular-variational-eigensolvers/tree/main}{public project repository}. The production variational-eigensolver implementation is a proprietary asset of Pattern Recognition Pty Ltd and is not publicly released. The deposited artifacts are sufficient to regenerate all reported tables, figures, and Hamiltonian-identity checks from the archived numerical outputs, but they do not provide an independent end-to-end rerun of the variational optimization.

\section*{Author Contributions}
Azadeh Alavi and Fatemeh Kouchmeshki contributed equally to this work.

Azadeh Alavi: conceptualisation, methodology, mathematical formulation, quantum-circuit design, benchmark design, formal analysis, interpretation of results, project administration, supervision, writing-original draft, and writing-review and editing.

Fatemeh Kouchmeshki: conceptualisation, methodology, software development, implementation, benchmarking, data curation, validation, formal analysis, interpretation of results, writing-original draft, and writing-review and editing.

Hossein Akhoundi: software development, implementation support, benchmarking support, validation, data curation, technical review, and writing-review and editing.

Abdolrahman Alavi: software development, implementation support, benchmarking support, validation, data curation, technical review, and writing-review and editing.

Muhammad Usman: Writing-review and editing.

Ke Deng: Writing-review and editing.

Yongli Ren: Writing-review and editing.

All authors reviewed and approved the final manuscript.

\section*{Competing Interests}

Azadeh Alavi is a Lecturer at RMIT University and a director of Pattern Recognition Pty Ltd. Fatemeh Kouchmeshki is affiliated with Pattern Recognition Pty Ltd. Pattern Recognition Pty Ltd may use publicly disclosed methods related to this work in independent research, prototyping, competition, and product-development activities. The authors declare that these interests have been disclosed and do not affect the scientific content, calculations, results, or conclusions of the manuscript.

\section*{Acknowledgment}
The authors acknowledge editorial drafting and consistency-checking assistance from OpenAI's ChatGPT and Anthropic's Claude. The authors take responsibility for the scientific content, calculations, and final manuscript decisions.

\bibliographystyle{IEEEtran}
\bibliography{references}

\end{document}